\documentclass[preprint,review,12pt]{elsarticle}

\usepackage{amsmath}
\usepackage{amssymb}
\usepackage{amsthm}

\usepackage{lineno}

\usepackage{bm}
\usepackage{epstopdf}
\usepackage{hyperref}
\usepackage{booktabs}
\usepackage{textcomp}
\usepackage{xcolor}
\usepackage{setspace}
\usepackage{geometry}
\usepackage{adjustbox}
\usepackage{rotating}
\usepackage{appendix}
\usepackage{physics}
\usepackage{subcaption}
\usepackage{multirow}

\biboptions{sort&compress}

\let\oldequation\equation
\let\oldendequation\endequation

\renewenvironment{equation}
  {\linenomathNonumbers\oldequation}
  {\oldendequation\endlinenomath}

\graphicspath{{Fig/}}

\journal{}

\newcommand{\bs}{\boldsymbol}
\newcommand{\wt}{\widetilde}

\newcommand{\etal}{\textit{et al}.}

\newcommand{\EQ}{\begin{equation}}
\newcommand{\EN}{\end{equation}}
\newcommand{\EQA}{\begin{eqnarray}}
\newcommand{\ENA}{\end{eqnarray}}

\newcommand{\lrr}[1]{\left(#1\right)}

\newcommand{\lra}[1]{\left\langle#1\right\rangle}

\begin{document}

\begin{frontmatter}



\title{Synergy of turbulence and thermo-diffusive effects on the intermittent boundary-layer flashback of swirling flames}


\author[fir]{Shiming Zhang}
\author[fir]{Zhen Lu}
\ead{zhen.lu@pku.edu.cn}
\author[fir,sec]{Yue Yang}

\address[fir]{State Key Laboratory for Turbulence and Complex Systems, College of Engineering, Peking University, Beijing 100871, China}
\address[sec]{HEDPS-CAPT, Peking University, Beijing 100871, China}

\begin{abstract}

We simulated the intermittent boundary-layer flashback (BLF) of hydrogen-enriched swirling flames using large-eddy simulation (LES) with the flame-surface-density (FSD) method. 
Three cases of intermittent BLF, characterized by periodic flame entry and exit of the mixing tube, are presented.
The intermittent BLF characteristics varied with the hydrogen volume fraction. 
Small flame bulges entered and exited the mixing tube in low hydrogen-enrichment cases. 
The duration of intermittent BLF events and BLF depth increased as the hydrogen content increased. 
Meanwhile, a large flame tongue penetrating deeply upstream characterised the highest hydrogen-enrichment case.
The mean BLF peak depths and standard deviations obtained through simulations aligned well with experimental data for low and moderate hydrogen-enrichment cases. 
However, LES-FSD underestimated the average BLF peak depth for the highest hydrogen-enrichment case.
Analysis of the flow-flame interaction revealed two mechanisms underlying the intermittent BLF phenomena.
The flame bulges' oscillation near the outlet is caused by the reverse flow induced by the recirculation zone. 
At the same time, the deep intermittent BLF occurrs due to the boundary layer separation induced by the large turbulent burning velocity, resulting from the synergy of turbulence and thermo-diffusive effects.

\end{abstract}


\end{frontmatter}



\section{Introduction\label{sec:introduction}}

Hydrogen-enriched fuels contribute to achieving zero-carbon and low-emission goals. 
However, the potential risk of boundary-layer flashback (BLF)~\cite{Lewis1943, Kalantari2017} increases with the hydrogen enrichment~\cite{Levinsky2021}. 
The intermittent BLF, characterized by the periodic protrusions of the flame front, is observed when the swirling flame is close to the BLF limit~\cite{Nauert2007, Heeger2010, Schneider2018, Schneider2020, Pancharia2022, Ebi2023}.
This dynamical intermittent BLF may serve as an early warning~\cite{Schneider2018, Schneider2020} of the extreme events~\cite{Hassanaly2021}. 

The BLF limit marks the transition from flame stabilization to flashback, and is defined by either a minimum bulk velocity or a maximum equivalence ratio required for flame stabilization in burners~\cite{Kalantari2017}.
The modelling of the BLF limit has been extensively studied~\cite{Lewis1943, Ebi2021, Hoferichter2017, Zhang2023}.
Lewis and von Elbe~\cite{Lewis1943} proposed the critical gradient model on the BLF limit of non-swirling flames.
Hoferichter~\etal~\cite{Hoferichter2017} developed a prediction model based on the boundary layer separation criterion.
Considering the effects of flame stretching and BLF modes, Zhang~\etal~\cite{Zhang2023} developed an algebraic model to predict the BLF limits, especially for the premixed hydrogen-enriched swirling flames at high pressures.
However, these models provide only a critical value and cannot be applied to the intermittent BLF condition.

Several experiments~\cite{Nauert2007, Heeger2010, Schneider2018, Schneider2020, Pancharia2022, Ebi2023} have reported the intermittent BLF near the flashback limits.
Nauert~\etal~\cite{Nauert2007} conducted experiments to classify different flame states during the rapid transition from stable flames to flashback. They observed that turbulence intensity and flow conditions strongly influence flashback.
Heeger~\etal~\cite{Heeger2010} identified small needle-shaped flame bulges submerged in the mixing tube that continuously rotate in the swirl direction.
Schneider and Steinberg~\cite{Schneider2018, Schneider2020} observed intermittent BLF through small protrusions and a large flame tongue at different hydrogen enrichment conditions. 
The depth and duration of intermittent BLF increased with higher equivalence ratios and hydrogen volume fractions.
Although the experiments have provided qualitative photographs and quantitative statistics on the depth and frequency of intermittent BLF, details of the flow fields remain necessary to understand the mechanism of intermittent BLF and its transition to complete BLF.

LES has emerged as a valuable numerical tool to study the extreme and rare events in combustion~\cite{Raman2019}, including BLF.
Several groups~\cite{Lietz2015, Jiang2021, Xia2022, Zhang2023, Zhang2024} have performed LES for the BLF of swirling flames at atmospheric and elevated pressures. 
The flame tongue structure in the BLF of swirling flames was simulated using the flamelet-based models~\cite{Lietz2015, Jiang2021} at different fuel mixtures and stratification conditions. 
Xia~\etal~\cite{Xia2022} investigated the effects of numerical boundary conditions on BLF using the artificial thickened flame model. 
Zhang~\etal~\cite{Zhang2023} employed the flame-surface-density (FSD) method to predict the BLF limits of hydrogen-enriched swirling flames. 
Zhang~\etal~\cite{Zhang2024} examined the effects of Soret and differential diffusion on the BLF process. 
Although the BLF of swirling flames has been simulated via LES, most existing numerical research has focused on the complete BLF or flame stabilization states, leaving the intermittent BLF relatively unexplored.

In the present work, we simulated the intermittent BLF of swirling flames with different levels of hydrogen enrichment using LES-FSD.
This approach has been validated for the planar turbulent premixed flames~\cite{Zhang2021} and BLF of swirling flames~\cite{Zhang2023}.
Our investigation focuses on the dynamics of intermittent BLF, an area that, to the authors' knowledge, lacks numerical studies. 
In the hydrogen and hydrogen-enriched turbulent flames, the synergy of turbulence and thermo-diffusive effects plays a crucial role~\cite{Lu2020, Lu2022, Berger2022, Su2022}. 
As the intermittent BLF characteristics change with hydrogen volume fraction, we aim to elucidate how this synergy affects the intermittent BLF process. 
By employing the LES-FSD approach, we captured the complex interactions between the flame and the flow field, providing insights into the mechanisms driving these intermittent events.

The rest of the paper is organized as follows.
The numerical settings are introduced in Sec.~\ref{sec:overview}, including the LES-FSD method and simulation setups.
The LES-FSD results on the intermittent BLF are presented and analyzed in Sec.~\ref{sec:results}.
Conclusions are summarized in Sec.~\ref{sec:conclusion}.

\section{Numerical model and simulation overview\label{sec:overview}}

\subsection{LES-FSD method\label{subsec:method}}

We simulated the swirling flames using LES-FSD, solving the governing equations on mass, momentum, and three scalars, including the progress variable $c$, generalized FSD $\Sigma$, and total enthalpy $h$~\cite{Zhang2021, Zhang2023}
\begin{equation}
	\frac{\partial\overline{\rho}}{\partial t}
	+ \nabla\cdot(\overline{\rho}\widetilde{\bs{u}})
	= 0,
	\label{eq:mass}
\end{equation}
\begin{equation}
	\frac{\partial(\overline{\rho}\widetilde{\bs{u}}) }{\partial t}
    + \nabla\cdot(\overline{\rho}\widetilde{\bs{u}}\widetilde{\bs{u}})
    + \nabla\cdot(\overline{\rho}\widetilde{\bs{u}\bs{u}}-\overline{\rho}\widetilde{\bs{u}}\widetilde{\bs{u}})
    =
    \nabla\cdot\overline{\bs{\tau}} - \nabla\overline{p},
    \label{eq:momentum}
\end{equation}
\begin{equation}
	\frac{\partial(\overline{\rho}\widetilde{c}) }{\partial t}
    + \nabla\cdot(\overline{\rho}\widetilde{\bs{u}}\widetilde{c})
    + \nabla\cdot(\overline{\rho}\widetilde{\bs{u}c}-\overline{\rho}\widetilde{\bs{u}}\widetilde{c})
    =
    \overline\rho \langle s_d \rangle_A  \Sigma,
    \label{eq:c}
\end{equation}
\begin{equation}
    \pdv{\Sigma}{t} + \nabla\cdot\lrr{\wt{\bs{u}}\Sigma}
    + \nabla\cdot\lrr{\lra{\bs{u}}_A-\wt{\bs{u}}}\Sigma
    =
    \mathcal{S} + \mathcal{P} + \mathcal{C},
    \label{eq:FSD}
\end{equation}
\begin{equation}
	\frac{\partial(\bar{\rho}\tilde{h}) }{\partial t}
    + \nabla\cdot(\overline{\rho}\widetilde{\bs{u}}\widetilde{h})
    + \nabla\cdot(\overline{\rho}\widetilde{\bs{u}h}-\overline{\rho}\widetilde{\bs{u}}\widetilde{h})
    =
    \nabla\cdot\overline{(\rho D \nabla h) }
    + {\overline{\mathcal{Q}}_{h}},
    \label{eq:enthalpy}
\end{equation}
where $t$, $\rho$, $\bs{u}$, $p$, $\bs{\tau}$, and $D$ are the time, density, velocity, pressure, viscous stress, and diffusivity, respectively;
$\overline{q}$, $\wt{q}$, and $\lra{q}_A$ denote the regular filtering, Favre filtering, and average over the flame surface, respectively;
The three terms $\mathcal{S}$, $\mathcal{P}$, and $\mathcal{C}$ on the right-hand side of the FSD equation, in Eq.~\eqref{eq:FSD}, model the straining, propagation, and curvature effects of the flame surface as
\begin{align}
    \mathcal{S}&  = \left(\nabla\cdot\widetilde{\bs{u}}-\bs{N}:\nabla\widetilde{\bs{u}} + { \Gamma \sqrt{k_{sgs}}}/{\hat{\Delta}}\right)\Sigma\\
    \mathcal{P}&  = - \nabla\cdot \left( \langle s_d \rangle_A \langle \bs{n} \rangle_A\Sigma\right)\\
    \mathcal{C}&  =  \langle s_d \rangle_A \langle \kappa \rangle_A \Sigma
	- {\alpha s_{L}^0(\wt{h}) \Sigma^2}/({1-\widetilde{c}})
\end{align}
where $\langle \bs{n} \rangle_A = - \nabla\wt{c}/\Sigma$ is the modelled surface-averaged normal vector,
$ \langle \kappa \rangle_A = \nabla\cdot\lra{\bs{n}}_A$,  
$\alpha = 1 - \lra{\bs{n}}_A\cdot\lra{\bs{n}}_A$, 
and $\bs{N} = \lra{\bs{n}}_A\lra{\bs{n}}_A + \frac{1}{3}\alpha\bs{I}$~\cite{Hawkes2000b},
$\hat{\Delta} = 5\Delta$ is the filter size~\cite{Boger1998} for scalars in combustion,
$\Delta$ is the filter size for velocity, 
$\Gamma$ is the efficiency function~\cite{Angelberger1998}, 
$k_{sgs}$ is the subgrid turbulent kinetic energy,
and $s_L^0(\wt{h})$ is the flame speed of a non-adiabatic one-dimensional freely propagating flame.

To account for the thermo-diffusive effects and the influence of the flame curvature in hydrogen-enriched flames, we modelled the surface-averaged local flame displacement speed $\lra{s_d}_A$ as~\cite{Zhang2021}
\EQ\label{eq:sd}
    \lra{s_d}_A = 
    \dfrac{\rho_{u} I_0 s_{L}^0 (\wt{h})}{\overline\rho}
	-
	D{\langle \kappa \rangle_A},
\EN
where $\rho_u$ denotes the density of the unburnt mixture and $I_0 = s_L(K, h)/s_L^0(h)$ is the stretch factor. 
The consumption speed of the stretched flame, $s_L(K, h)$, is obtained based on the non-adiabatic counterflow flames at various Karlovitz factor $K=a_t \delta_L^0(h) / s_L^0(h)$, where $a_t$ is the strain rate and $\delta_L^0(h)$ is the thermal thickness of the freely propagating flame.

We simulated the non-adiabatic one-dimensional freely propagating flames and stretched flames to build up the databases on $s_L^0$ and $I_0$.
The heat transfer effects were modelled via a modified energy equation~\cite{Proch2015} in the simulation of non-adiabatic flames.
In LES-FSD, $s_{L}^0$ and $I_0$ are retrieved from the look-up table using the enthaply $\wt{h}$ and Karlovitz factor $K=\mathcal{S}\delta_L^0(\wt{h})/[\Sigma s_L^0(\wt{h})
]$ in the flow field~\cite{Zhang2023}.
The model on surface-averaged local displacement speed in Eq.~\eqref{eq:sd} considers the thermo-diffusive effects and flame surface curvature. 
The enthalpy $\wt{h}$ is transported to account for heat loss and super-adiabatic effects. 
The model's improvements for lean hydrogen and hydrogen-enriched premixed flames have been validated on the turbulent planar~\cite{Zhang2021} and swirling~\cite{Zhang2023} flames.

\subsection{Simulation setup\label{subsec:setup}}

\begin{figure}[ht]
	\centering
	\includegraphics[width=88 mm]{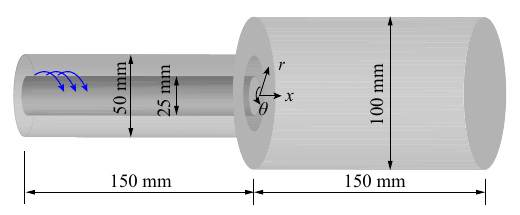}
	\caption{
		Schematic of the computational domain for LES-FSD, where the blue arrows denote the swirling inflow direction.
	}
	\label{fig:config}
\end{figure}

We simulated the intermittent BLF of hydrogen-enriched flames in a swirling burner with a central bluff body~\cite{Schneider2020}. 
The burner consists of a mixing tube and a combustion chamber, as shown in Fig.~\ref{fig:config}. 
The mixing tube has an outer diameter of 50 mm and a length of 150 mm, while the combustion chamber measures 100 mm in diameter and 150 mm in length. 
A central bluff body, 25 mm in diameter, is positioned within the mixing tube.
In the cylindrical coordinate, we set the centre of the mixing tube outlet as $x=0$, as marked in Fig.~\ref{fig:config}.

The fuel and air mixture is fed through the inlet at 297 K and atmospheric pressure, with a swirl number $S=0.9$. 
The air and methane flow rates remain constant across cases at 720 and 35.5 standard litres per minute, respectively.
We conducted three intermittent BLF cases with different levels of hydrogen enrichment, as detailed in Table~\ref{tab:sim_cases}:
Case G1: $X_\mathrm{H_2}=42.3\%$, 
Case G2: $X_\mathrm{H_2}=47.2\%$, and
Case G3: $X_\mathrm{H_2}=51.3\%$.
These variations in hydrogen content result in different equivalence ratios $\phi$ and inlet bulk velocities $U_x$.
\begin{table}[htbp]
    \centering
	\caption{Operating conditions in the LES-FSD for the intermittent BLF of CH$_4$/H$_2$/air swirling flames.}
	\begin{tabular}{cccc}
		\hline
		Case  & $X_\mathrm{H_2}\;(\%)$ & $\phi$ & $U_{x}\;(\mathrm{m/s})$    \\
		\hline
        G1    & 42.3	& 0.556    & 8.845  \\
        G2    & 47.2	& 0.575    & 8.910  \\	
        G3    & 51.3    & 0.594    & 8.975  \\
		\hline
	\end{tabular}
	\label{tab:sim_cases}
\end{table}

A separate LES was conducted on a periodic mixing tube to establish the inlet velocity profile. 
To obtain a stable swirl number and turbulent intensity, we utilized a linear forcing method~\cite{Pierce1998, Carroll2013}, resulting in a ratio of inlet turbulence intensity to the mean streamwise velocity of 30\%. 
To minimize the impact of the inlet boundary condition on the combustion LES, the swirling inflow LES duration was over 1 second once a statistically stationary state was reached.

The simulations of intermittent BLF were initialized as cold flows with $\wt{c}=0$. 
Once the flow field was fully developed, we ignited the combustor by imposing a spherical flame at $x=$ 25 mm. 
Following flame propagation, stabilized swirling flames were obtained in the combustion chamber. 
As the hydrogen flow rate increased, distinct intermittent BLF was triggered.

The transport equations in LES-FSD were solved using the NGA code~\cite{Desjardins2008}. 
The momentum equations were discretized using a second-order, centered, kinetic-energy conservative scheme. 
Convection terms in the scalar transport equations were handled using a third-order weighted essentially non-oscillatory scheme~\cite{Liu1994}. 
A semi-implicit Crank--Nicolson scheme~\cite{Pierce2004} was employed for the time marching of the transport equations. 
The dynamic Smagorinsky model~\cite{Pierce2004} closed the subgrid stresses, turbulent kinetic energy, and scalar fluxes. 
Further details about the implementation can be found in Refs.~\cite{Zhang2021, Zhang2023}. 

\begin{figure}[ht]
	\centering
	\includegraphics[width=0.5\textwidth]{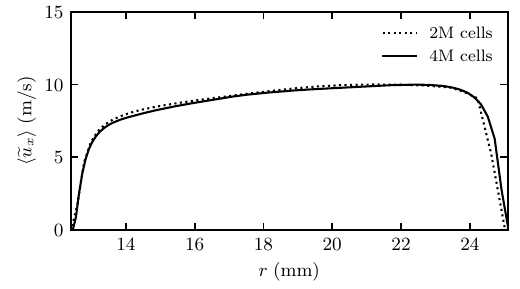}
	\caption{Radial profiles of averaged velocity $\widetilde{u}_x$ at $x=-20$ mm near the exit of mixing tube obtained from LES with 2 million cells (dotted line) and 4 million cells (solid line).}
	\label{fig:mesh}
\end{figure}

We discretized the computational domain with 2 million cells. 
In the mixing tube, we refined the near-wall mesh to ensure 15 grid points within 30 non-dimensional wall distance units $y^+$. 
Time stepping was selected to maintain a Courant--Friedrichs--Lewy number below 0.5, ensuring numerical stability and accuracy.
A mesh sensitivity assessment was performed on case G3, utilizing two grids comprising 2 million and 4 million cells.
Results demonstrate good mesh convergence on the velocity profile, with the averaged velocity near the exit of the mixing tube shown in Fig.~\ref{fig:mesh}.

\section{Results\label{sec:results}}

\subsection{Intermittent BLF\label{subsec:IBLF}}

We conducted LES-FSD on the swirling flames with different levels of hydrogen enrichment.
Our LES-FSD obtained distinct BLF phenomena as the hydrogen flow rate increased. 
At low and moderate hydrogen flow rates, small flame bulges intermittently entered and exited the upstream mixing tube. 
At a high hydrogen flow rate, the swirling flame exhibited a large-scale flame tongue structure. 
The flame tongue alternated between advancing, holding position, and retreating within the mixing tube multiple times before fully retreating into the combustion chamber. 
We classify these two phenomena as intermittent BLF, characterized by the periodic flame entry and exit at the mixing tube outlet. 
Further increases in the hydrogen flow rate resulted in continuous BLF along the central bluff body, representing a complete BLF phenomenon. 

\begin{figure}[ht]
	\centering
	\includegraphics[width=\textwidth]{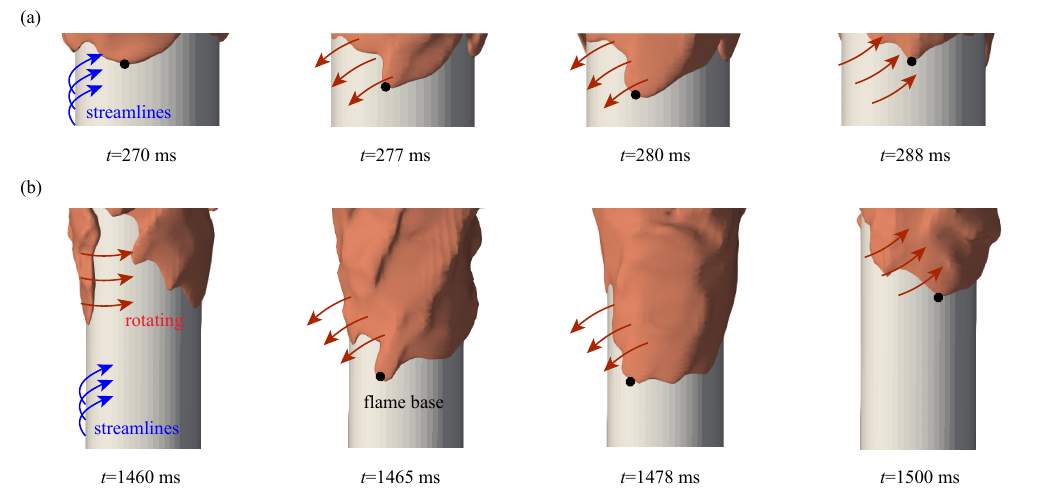}
	\caption{
     Flame surfaces obtained via LES-FSD at different times in case (a) G1 and (b) G3.
     The red isosurfaces correspond to $\wt{c}=$0.81 and 0.77 in cases G1 and G3, respectively. 
     The blue and red arrows indicate the swirling flow and flame propagation directions, respectively.
	}
	\label{fig:flame}
\end{figure}

Figure~\ref{fig:flame} illustrates the two intermittent BLF processes observed in cases G1 and G3. 
The flame surfaces obtained in LES-FSD are depicted by red isosurfaces, corresponding to $\wt{c}=$0.81 and 0.77 in cases G1 and G3, respectively. 
These values represent the locations of maximum heat release rate in the laminar flames. 
The central bluff body in the mixing tube is denoted by a white solid surface.
The blue and red arrows indicate the swirling flow and flame propagation directions, respectively. 
We define the flame base as the lowest position of the flame surface~\cite{Zhang2023}, marked by black points in Fig.~\ref{fig:flame}. 
The axial location of the flame base is referred to as the BLF depth, providing a quantitative measure of flame penetration into the mixing tube.

In case G1, characterized by a low level of hydrogen enrichment, small flame bulges periodically propagated into the mixing tube and retreated into the combustion chamber. 
For instance, between $t=$ 277 and 280 ms, a flame bulge propagated against the swirling flow when entering the mixing tube. 
Subsequently, it retreated into the combustion chamber in the same direction as the swirling flow from $t=$ 280 to 288 ms. 
Case G2 exhibited flame propagation patterns similar to case G1, with flames advancing as small-scale flame bulges. However, the propagation distance and size of the flame bulges in G2 exceeded those observed in G1, indicating a more pronounced intermittent BLF behaviour with increased hydrogen enrichment.

Case G3, characterized by a high hydrogen flow rate, exhibited a large-scale flame tongue similar to the complete BLF condition~\cite{Zhang2023}. 
In this case, the flame tongue frequently oscillated upstream and downstream within the mixing tube, rotating in the same direction as the swirling flow. 
The BLF depth generally exceeded that observed in cases with small flame bulges.

When the flame tongue advanced further upstream, it propagated against the incoming swirling flow before retreating. 
The advancing and retreating directions of the large-scale flame tongue mirrored those of the small protrusions case. 
Note that there is no clear indication of when the flame tongue will propagate upstream. 
Instead, it may alternate between holding its position, advancing, and retreating several times before completely retreating into the combustion chamber.

\begin{figure}[ht!]
	\centering
    \begin{subfigure}{0.9\textwidth}
        \centering
	    \includegraphics[width=0.5\textwidth]{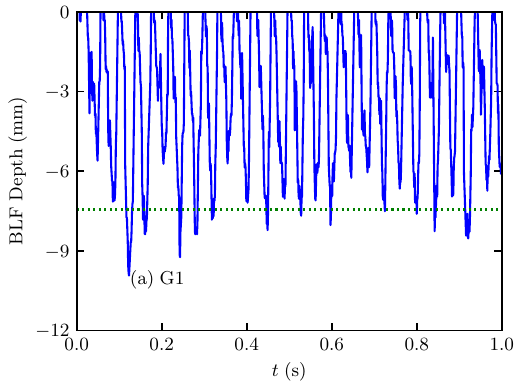}
    \end{subfigure}
    \begin{subfigure}{0.9\textwidth}
        \centering
	    \includegraphics[width=0.5\textwidth]{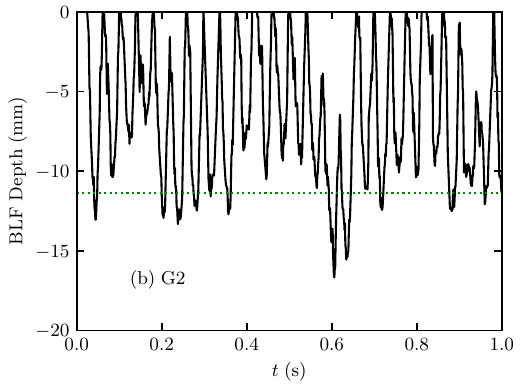}
    \end{subfigure}
    \begin{subfigure}{0.9\textwidth}
        \centering
	    \includegraphics[width=0.5\textwidth]{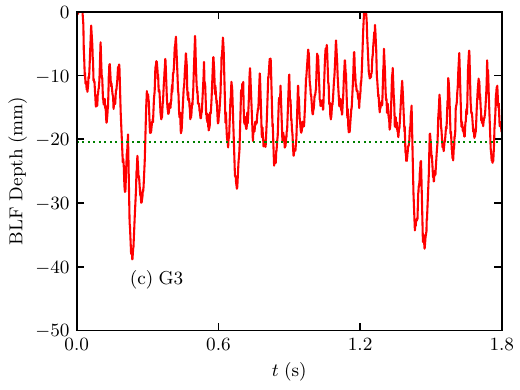}
    \end{subfigure}
	\caption{Temporal variations of BLF depth in cases (a)  G1, (b) G2, and (c) G3, where the green dotted line represents the mean BLF peak depth.}
	\label{fig:depth}
\end{figure}

Figure~\ref{fig:depth} illustrates the temporal variation of the BLF depth in cases G1, G2, and G3.
In case G1, the maximum BLF depth of each event remained below 10 mm. 
In contrast, case G3 exhibited more pronounced fluctuation in BLF depth, reflecting the dynamic behaviour of the large-scale flame tongue. 
The BLF depth in case G3 generally exceeded that of cases with smaller flame bulges, as shown in Fig.~\ref{fig:depth}c. 
The flame tongue in case G3 exhibited diverse behaviours: it propagated upstream, as observed at $t=$ 0.2 s; maintained its position, as seen during the interval from $t=$ 0.4 to 1.1 s; or retreated to the mixing tube outlet, as demonstrated at $t=$ 1.2 s. 
Case G2 represented an intermediate scenario between G1 and G3.
Intermittent BLF events to greater depths occasionally occured, as evidenced by a BLF depth exceeding 15 mm at $t=$ 0.6 s, depicted in Fig.~\ref{fig:depth}b. 

Figure~\ref{fig:depth} shows many local BLF depth maxima, encompassing both intermittent BLF event peaks and minor oscillations. 
To filter oscillations and obtain each BLF peak depth, we employed the identical peak-seeking approach used in the experiments~\cite{Schneider2020}. 
The dotted lines in Fig.~\ref{fig:depth} represent the mean BLF peak depths for each case. 
Table~\ref{tab:depth} summarizes the mean BLF peak depth acquired in LES-FSD and experiments for cases G1, G2, and G3.
As the hydrogen flow rate increased, case G2 exhibited a slight rise in BLF peak depth compared to case G1.
The LES-FSD results aligned well with the experimental data for cases G1 and G2. 
Case G3, characterized by the highest hydrogen volume fraction, exhibited intermittent BLF at greater BLF depths. 
Although our simulations captured the dynamics qualitatively, LES-FSD underestimated the mean BLF peak depth in case G3. 
The discrepancy may be due to the imperfection in modelling, particularly in flame-wall interactions and thermo-acoustic effects.
\begin{table}[htbp]
    \centering
    \caption{Mean BLF peak depths obtained in LES-FSD and experiments of cases G1, G2, and G3.}
	\begin{tabular}{cccc}
		\hline
        \multirow{2}{*}{Case} & \multirow{2}{*}{$X_\mathrm{H_2}\;(\%)$} & \multicolumn{2}{c}{Mean BLF peak depth (mm)} \\
		& & Exp. & LES-FSD    \\
		\hline
        G1	& 42.3	& -8.3  & -7.5  \\
        G2	& 47.2	& -9.5  & -11.4  \\	
        G3	& 51.3	& -60.6 & -20.5  \\
		\hline
	\end{tabular}
	\label{tab:depth}
\end{table}

\subsection{Flow-flame interactions\label{subsec:effects}}

Two types of intermittent BLF phenomena were observed in both simulations and experiments: small flame bulges oscillating near the mixing tube outlet and a large-scale flame tongue penetrating deep upstream. 
As the hydrogen enrichment and equivalence ratio increase from case G1 to G3, 
the laminar flame speed rises, consequently increasing the turbulent burning velocity. 
However, this increase in flame speed does not fully explain the different BLF behaviours observed.
To elucidate the underlying mechanism of different intermittent BLF phenomena, we analyze the flow-flame interaction during the intermittent BLF process in the following.

\begin{figure}[ht!]
	\centering
    \begin{subfigure}{.9\textwidth}
        \centering
	    \includegraphics[width=0.5\textwidth]{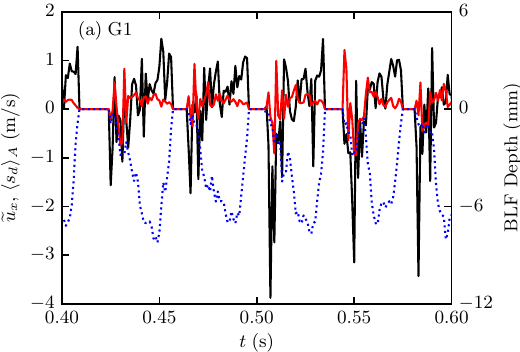}
    \end{subfigure}
    \begin{subfigure}{.9\textwidth}
        \centering
	    \includegraphics[width=0.5\textwidth]{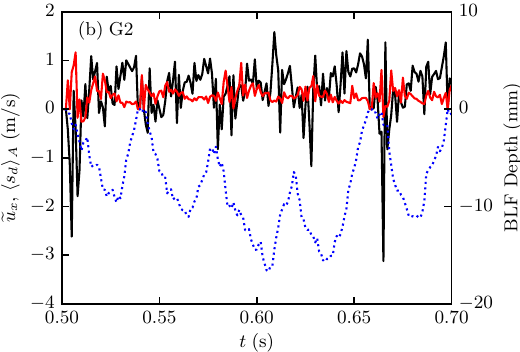}
    \end{subfigure}
    \begin{subfigure}{.9\textwidth}
        \centering
	    \includegraphics[width=0.5\textwidth]{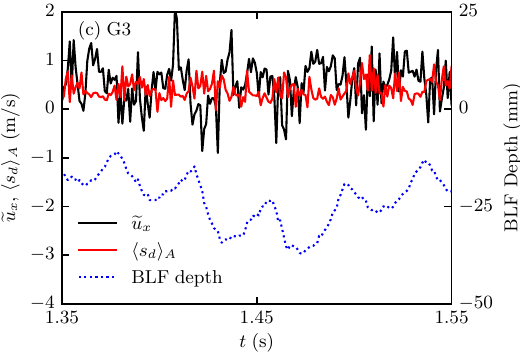}
    \end{subfigure}
	\caption{Flow velocity $\wt{u}_x$ (black solid line), local flame displacement speed $\lra{ s_d }_A$ (red solid line), and BLF depth (blue dotted line) at the flame base in cases (a)  G1, (b) G2, and (c) G3.}
	\label{fig:speed_t}
\end{figure}

Figure~\ref{fig:speed_t} shows the axial flow velocity $\wt{u}_x$, local flame displacement speed $\lra{ s_d }_A$, and BLF depth at the flame base in cases G1, G2, and G3. 
Each case is presented over a period of 0.2 s. 
When the flame stabilized without penetration into the mixing tube, multiple flame fronts appeared at $x=$ 0. 
In these instances, we plot $\wt{u}_x$ and $\lra{ s_d }_A$ as 0 when BLF depth equals 0.
This simplification does not affect our conclusions, as our analysis focuses on the flow-flame interaction during the intermittent BLF within the mixing tube.

Each intermittent BLF event begins with a BLF depth of 0. 
The onset of intermittent BLF events at the mixing tube outlet is characterized by large negative $\wt{u}_x$ values.
These large negative $\wt{u}_x$ are evident for every intermittent BLF event in Fig.~\ref{fig:speed_t}a for case G1 and the events at $t=$ 0.50 and 0.66 s in Fig.~\ref{fig:speed_t}b for case G2.
Similar observations are made at $t=$ 0.0 and 1.2 s for case G3, although not shown in Fig.~\ref{fig:speed_t}. 
In contrast, the intermittent BLF within the mixing tube initiates with a smaller negative velocity, as observed in cases G2 and G3.

\begin{figure}[ht]
	\centering
    \includegraphics[width=0.45\textwidth]{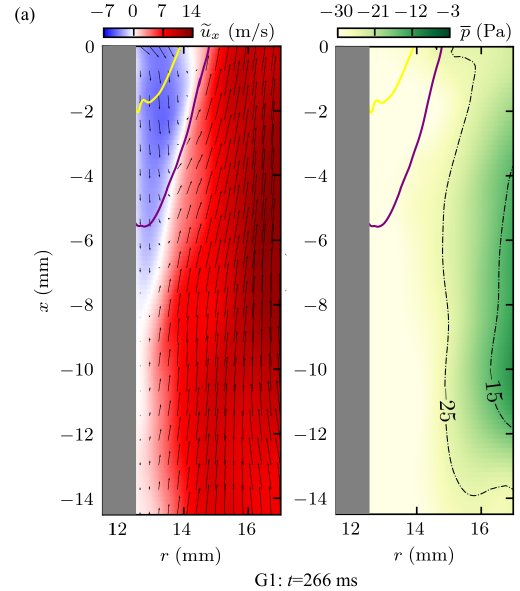}
    \includegraphics[width=0.45\textwidth]{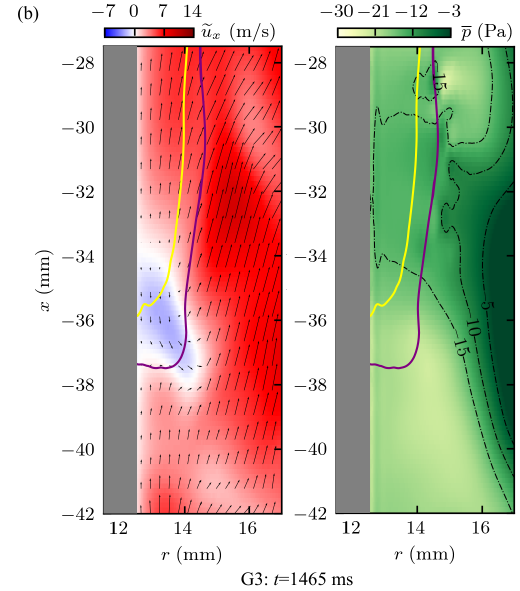}
	\caption{Instantaneous contours of $\wt{u}_x$ on the $x$-$r$ surface in case (a) G1 and (b) G3, and corresponding contour lines of $\wt{c}=$0.1 (purple) and $\wt{c}$ (yellow) with the maximum heat release rate in the laminar flame.}
	\label{fig:ux_cloud}
\end{figure}

The association between the negative flow velocity and the onset of intermittent BLF events warrants examination of the flow fields around the flame base in the two types of intermittent BLF phenomena. 
Figure~\ref{fig:ux_cloud} presents the instantaneous contours of $\wt{u}_x$ and $\bar{p}$ on the $x$-$r$ surface for cases G1 and G3 at $t=$ 0.266 and 1.465 s, respectively.
Additionally, Fig~\ref{fig:ux_cloud} includes contour lines of the progress variable $\wt{c}$ at 0.1 and the maximum heat release rate location in the corresponding laminar flames for each case. 

For case G1 shown in Fig.~\ref{fig:ux_cloud}a, recirculation in the combustion chamber generates a reverse flow near the mixing tube outlet, propelling flames fronts upstream. 
No apparent pressure gradient exists in the boundary layer. 
As the reverse flow dissipates, the incoming flow pushes the flame fronts out of the mixing tube. 
For case G3 in Figure~\ref{fig:ux_cloud}b, the flame tongue is located at $x=$ -38 mm, far from the mixing tube outlet, at $t=$ 1.465 s. 
A reverse flow region is observed near the flame front, coupled with the reverse pressure gradient in the boundary layer. 
This reverse pressure gradient induces the boundary-layer separation, leading to the BLF in the mixing tube. 
This phenomenon is similar to the complete BLF process~\cite{Zhang2023}. 
The velocity and pressure profiles shown in Figs.~\ref{fig:speed_t} and \ref{fig:ux_cloud} suggest that the intermittent BLF around the mixing tube outlet and within the mixing tube are caused by the recirculation and boundary-layer separation, respectively.

\begin{figure}[ht]
	\centering
	\includegraphics[width=0.45\textwidth]{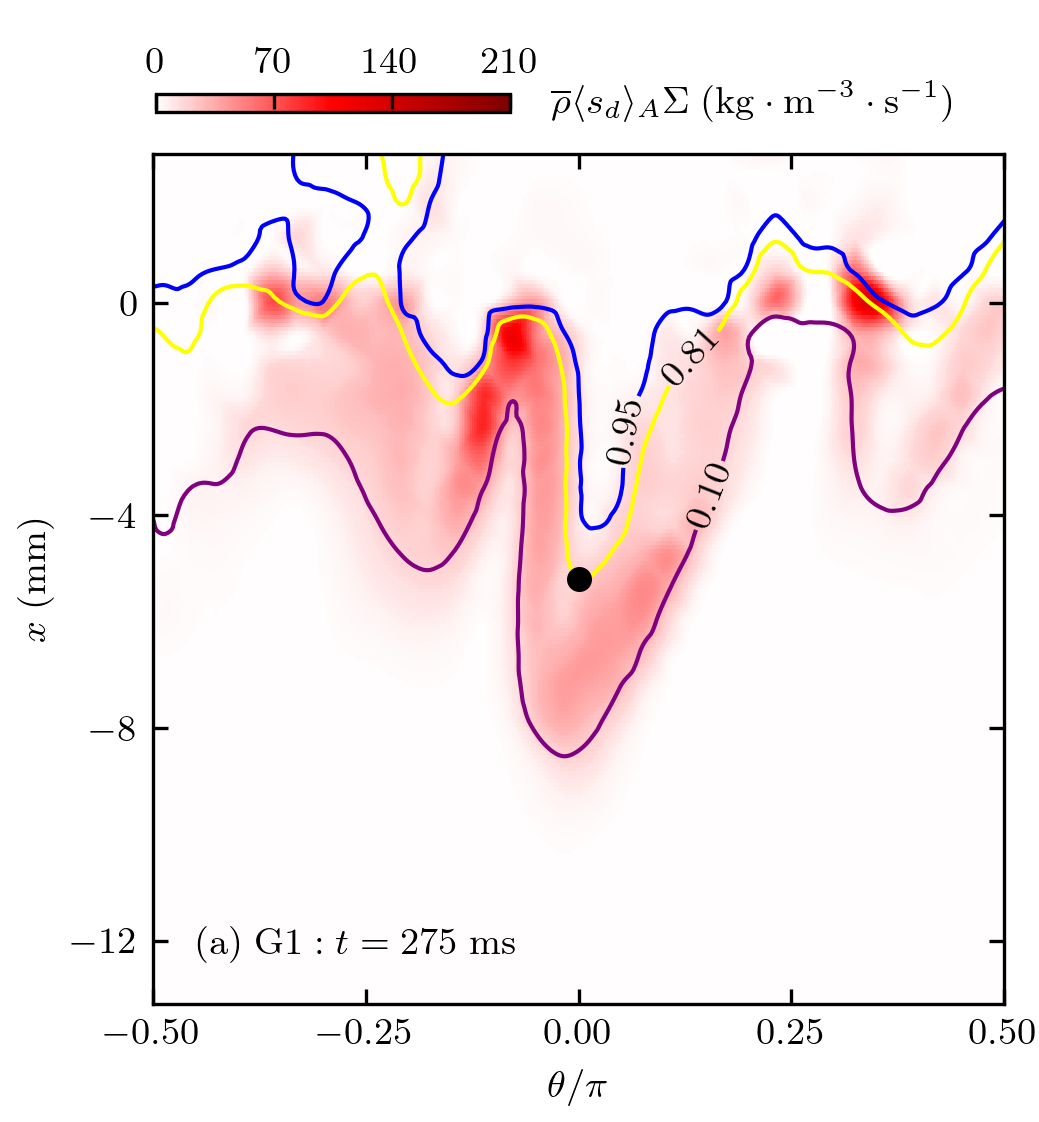}
	\includegraphics[width=0.45\textwidth]{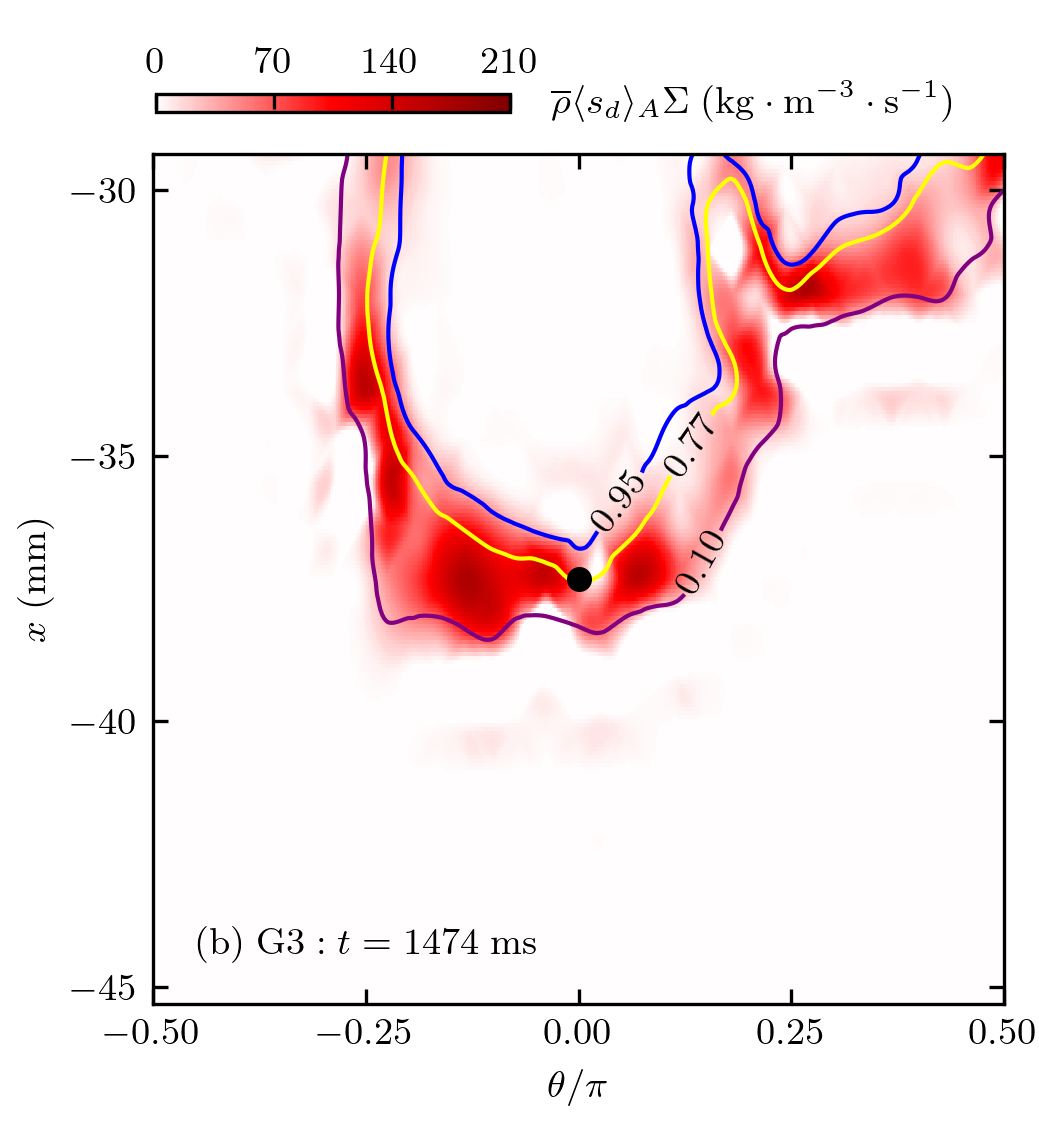}
	\caption{Instantaneous contours of the source term $\overline\rho \langle s_d \rangle_A  \Sigma$ of Eq.~\eqref{eq:c} for cases (a) G1 and (b) G3. The solid  lines denote different $\widetilde{c}$ contours.}
	\label{fig:fsd_contours}
\end{figure}

As shown in experiments and simulations~\cite{Hoferichter2017, Zhang2023}, the boundary-layer separation is induced by the flame propagation-induced pressure rise. 
Figure~\ref{fig:fsd_contours} plots the instantaneous contours of the source term $\overline\rho \langle s_d \rangle_A \Sigma$ of the progress variable transport equation for cases G1 and G3 on the $x$-$\theta$ surface. 
The source term $\overline\rho \langle s_d \rangle_A \Sigma$ is equivalent to the mass flux across the turbulent flame front. 
It is apparent that case G3 has larger turbulent burning velocity near the flame front.
Consequently, the large turbulent burning velocity leads to a reverse pressure gradient along the boundary layer, inducing the reverse flow and BLF in the mixing tube.

\begin{figure}[ht]
    \centering
	\includegraphics[width=0.45\textwidth]{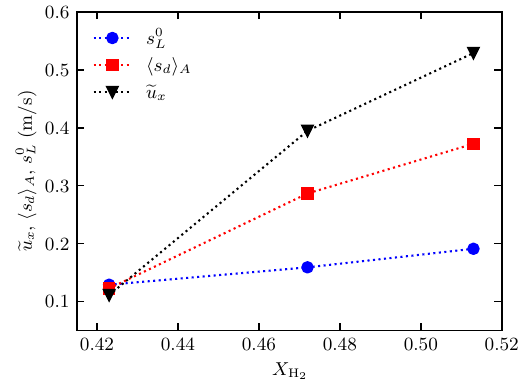}
    \includegraphics[width=0.45\textwidth]{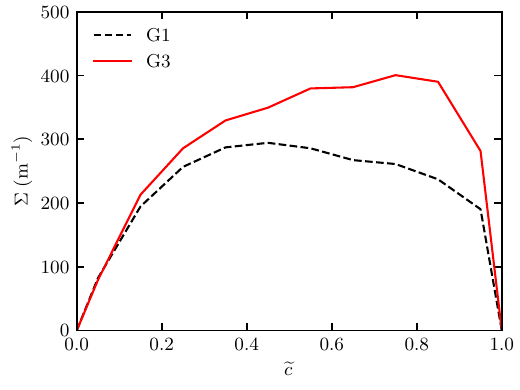}
    \caption{
    (a) Comparisons of the time-averaged $\langle s_d \rangle_A$ and $\widetilde{u}_x$ near the flame base and the laminar flame speed $s_L^0$ in cases G1, G2, and G3.
    (b) FSD conditional upon $\widetilde{c}$ around the flame base of cases G1 and G3.
    }
    \label{fig:local_stats}
\end{figure}

The larger turbulent burning velocity in case G3 is attributed to the synergy of turbulent flows and the thermo-diffusive effects of hydrogen-enriched flame~\cite{Lu2020, Lu2022, Berger2022}. 
Figure~\ref{fig:local_stats} presents flame data collected on the $x$-$r$ surface crossing the flame base and located within 5 mm of it.
From case G1 to G3, the average local flame speed rises threefold, despite only a slight increase in the laminar flame speed. 
This significant increase occurs because turbulent swirling flows strongly stretch and curve the flame surface. 
Consequently, the synergy of turbulence and thermo-diffusive effects increases the local flame speed, further enhancing the wrinkling of the turbulent flame surface.

Furthermore, the synergy between turbulence and thermo-diffusive effects introduces significant fluctuations in the turbulent burning velocity. 
The one standard deviation range exceeds the average flame speed, precluding its representation in Fig.~\ref{fig:local_stats}. 
Consequently, the reverse pressure gradient experiences similar fluctuations to the turbulent burning velocity, resulting in oscillations of the flame tongue within the mixing tube.

\section{Conclusion}\label{sec:conclusion}

In this study, we conducted LES using the FSD method to investigate the intermittent BLF in hydrogen-enriched swirling flames. 
Our simulations revealed distinct intermittent BLF characteristics across different levels of hydrogen enrichment, which aligned well with experimental observations.

The LES-FSD method, validated on turbulent premixed and swirling flames, effectively captured the dynamics of intermittent BLF. 
At low and moderate hydrogen enrichment, small flame bulges intermittently entered and exited the mixing tube. 
As hydrogen enrichment increased, these bulges penetrated deeper, and the duration of intermittent BLF events lengthened. 
In the highest hydrogen-enrichment case, BLF was dominated by a large-scale flame tongue that rotated, propagated, and retreated within the mixing tube, accompanied by sudden and deep BLF events.
Quantitatively, LES-FSD results were consistent with experimental data on mean BLF peak depths for low and moderate hydrogen-enrichment cases. 
However, for the highest hydrogen-enrichment case, the LES-FSD method underestimated the average BLF depth, indicating areas for further refinement in the simulation model. 

Our analysis of the coupling between the flow field and flames shows that the oscillation of flame bulges is primarily driven by the negative axial velocity region caused by flow recirculation at the outlet of the mixing tube. 
Meanwhile, case G3 exhibited a significantly higher turbulent burning velocity than case G1, attributed to the synergistic effects of turbulent swirling flows and the thermo-diffusive properties of hydrogen-enriched flames.
The large turbulent burning velocity induces the reverse pressure gradient in the boundary layer, leading to BLF in the mixing tube. 
Furthermore, the interaction between turbulence and thermo-diffusive effects introduces substantial fluctuations in the turbulent burning velocity. 
These fluctuations, in turn, caused oscillations in the reverse pressure gradient and the flame tongue within the mixing tube.

Note that higher oscillations of the intermittent BLF depth are filtered due to the limitations of LES. 
Additionally, our simulations do not account for thermo-acoustic effects, which could potentially influence sudden and deep BLF events.
Incorporating thermo-acoustic effects into the simulations may provide a more comprehensive understanding of the factors influencing BLF events. 
Furthermore, experimental validation at varying pressures and hydrogen concentrations would further enhance the robustness of the simulation models.

\section*{Acknowledgement}
We gratefully acknowledge Caltech, the University of Colorado at Boulder, and Stanford University for licensing the NGA code used in this work.
Numerical simulations were carried out on the Tianhe-2A supercomputer in Guangzhou, China.
This work has been supported in part by the National Natural Science Foundation of China (Grant Nos.~52306126, 22350710788, 11925201, 11988102, and 92270203), the National Key R\&D Program of China (No.~2020YFE0204200), and the Xplore Prize.


%
%

\bibliographystyle{elsarticle-num}
\bibliography{intermittentBLF.bib}





\end{document}